\documentclass[openacc]{rstransa}

\usepackage{multicol}
\usepackage{multirow}
\usepackage{array}

\usepackage[labelformat=simple]{subcaption}		

\begin{document}

\title{Robustness of convolutional neural networks to physiological ECG noise}

\author{
J. Venton$^{1}$, P. M. Harris$^{1}$, A. Sundar$^{1}$, N. A. S. Smith$^{1}$ and P. J. Aston$^{2,1}$}

\address{$^{1}$Department of Data Science, National Physical Laboratory, Teddington, UK\\
$^{2}$Department of Mathematics, University of Surrey, Guildford, UK\\}

\subject{machine learning, cardiovascular conditions, ECG}


\corres{Jenny Venton\\
\email{jenny.venton@npl.co.uk}}

\begin{abstract}
The electrocardiogram (ECG) is one of the most widespread diagnostic tools in healthcare and supports the diagnosis of cardiovascular disorders. Deep learning methods are a successful and popular technique to detect indications of disorders from an ECG signal. However, there are open questions around the robustness of these methods to various factors, including physiological ECG noise. In this study we generate clean and noisy versions of an ECG dataset before applying Symmetric Projection Attractor Reconstruction (SPAR) and scalogram image transformations. A pretrained convolutional neural network is trained using transfer learning to classify these image transforms. For the clean ECG dataset, F1 scores for SPAR attractor and scalogram transforms were 0.70 and 0.79, respectively, and the scores decreased by less than \(0.05\) for the noisy ECG datasets. Notably, when the network trained on clean data was used to classify the noisy datasets, performance decreases of up to \(0.18\) in F1 scores were seen. However, when the network trained on the noisy data was used to classify the clean dataset, the performance decrease was less than \(0.05\). We conclude that physiological ECG noise impacts classification using deep learning methods and careful consideration should be given to the inclusion of noisy ECG signals in the training data when developing supervised networks for ECG classification.
\end{abstract}


\begin{fmtext}

\end{fmtext}


\maketitle

\section{Introduction}
\subsection{Deep learning and physiological ECG signal noise}
Electrocardiogram (ECG) signals have long been used to support the diagnosis of cardiovascular disorders. Deep learning methods show encouraging results in ECG classification tasks and have recently seen a rapid increase in popularity \cite{Hong2020}. Noise and interference on the ECG signal are established causes of error in ECG diagnosis and interpretation \cite{Luo2010} and have been noted to affect both manual (clinician) and automated (machine learning) detection of ECG abnormalities \cite{Ribeiro2020}. A desirable property of a deep network is that the performance of the network is robust to perturbations in the input data. Network robustness to ECG noise of deep learning methods used to detect cardiovascular disorders is not well understood and there have been no studies directly addressing the issue. Sources of noise that degrade the quality of a dataset include both label noise (in terms of mislabelled data) and ECG signal noise (in terms of physiological noise on the signal). Here we focus on the impact of ECG signal noise, to gain an understanding of how physiological ECG noise impacts the robustness of deep learning methods.

\subsection{Transfer learning with deep networks}
While custom network architectures can be developed and trained from scratch to classify ECG signals \cite{Hong2020}, transfer learning is a popular method for utilising pretrained deep networks with new data. Transfer learning refers to the retraining of a pretrained network, for example a network pretrained using ImageNet \cite{ImageNet} data can be retrained using ECG data to classify ECG data. This training method is useful when there is a lack of data, computational resources or time, or to prototype models and carry out exploratory analysis.

ECG datasets often contain fewer than the large number of samples required to train a deep network from scratch, and in this case transfer learning is an attractive option. Furthermore, many well known network architectures have a demonstrated record of high performance. The focus of this study is to evaluate the robustness of deep learning to physiological ECG noise, rather than to optimise ECG classification performance of a custom architecture. Using an established pretrained network provides a solid foundation for this focus.

Convolutional neural networks (CNNs) are a class of deep network that is widely used for image classification and, alongside recurrent neural networks (RNNs), are commonly used for ECG classification \cite{Hong2020}. Both 1D CNNs applied to the raw ECG signal and 2D CNNs applied to ECG image transforms have been used to detect cardiovascular disorders from the ECG signal.

\subsection{Detecting cardiovascular disorders from the ECG signal}
Extensive work has been carried out to develop methods, including deep networks, that extract information from an ECG signal to support clinical decision making \cite{Hong2020,Lyon2017}. ECG image transformations are methods that convert a 1D ECG signal to a 2D image which can then be passed to a 2D CNN for training and classification.

The use of ECG image transforms allows both the use of 2D CNNs pretrained on the popular image dataset ImageNet \cite{Deng2009}, and the exploration of the impact of these image transforms on network robustness to noise. ECG image transforms capture frequency domain or morphology information that describe the underlying signal. Existing ECG image transforms and their applications include: the continuous wavelet transform (scalogram) for biometrics \cite{Byeon2019}, gray-level co-occurence matrix for morphological arrhythmia detection \cite{Sun2019}, recurrence plot to classify arrhythmias \cite{Mathunjwa2021}, distance distribution matrix to identify congestive heart failure \cite{Li2018} and the Symmetric Projection Attractor Reconstruction (SPAR) method for genetic mutation detection \cite{Aston2019}.

Networks trained to classify ECG image transforms are less common than networks trained to classify the ECG signal directly and their utility for pathology classification is still being explored. In particular, the impact of using ECG image transforms on network robustness is unclear.

\subsection{Objectives}
The main objectives of this study are to:
\begin{enumerate}
    \item Study the impact of the inclusion of physiological ECG noise in the input data on classification performance of a CNN (the robustness);
    \item Assess the impact of the inclusion of physiological noise in the training data on the robustness of a CNN;
    \item Determine whether different ECG image transforms increase or decrease robustness to different noise types.
\end{enumerate}

\section{Methods}
\subsection{ECG dataset}\label{sec:ECGdata}
Twelve lead ECG signals from the first source of data made available for the PhysioNet/Computing in Cardiology Challenge 2020 \cite{PerezAlday2020} were used, and all data used were open access. All subjects with one of the following three diagnostic classes were selected: atrial fibrillation (AF), healthy (Normal) and ST depression (STD). There were 2678 subjects in total: 976 AF, 918 Normal and 784 STD. Data were recorded at 500 Hz and signals were 8 to 138 seconds long.

Beyond selecting all subjects with one of the three chosen diagnostic class labels there were no further selection criteria applied. All data with precisely one of the three chosen class labels were used. Although the choice of classes was not clinically motivated, there were several motivations behind it. Firstly, to provide balanced class sizes since although addressing the impact of imbalanced class sizes on deep learning methods is important, as is developing models that can assign more than one class label, the focus of this study was to facilitate evaluation of robustness. Secondly, to reduce the number of classes the model had to identify, as a simpler model allowed a more thorough investigation of robustness. Finally, the healthy `Normal' category was chosen alongside two pathology classes to study any differences in robustness to physiological noise between healthy and pathological ECG signals.

Furthermore, as ECG image transforms were used, one ECG rhythm based class (AF) and one ECG morphology based class (STD) were chosen to assess any differing performance with different ECG image transforms. AF is a heart condition resulting in irregular heart rhythm and is characterised by small irregular \textit{f} (fibrillatory) waves on the ECG \cite{Goldberger2018}. STD is a morphological feature of an ECG signal and can be indicative of several conditions \cite{Pollehn2002}.

\subsection{Physiological ECG signal noise datasets}
To explore the robustness of CNNs to physiological ECG signal noise, the raw ECG dataset described in Section \ref{sec:ECGdata} was filtered to remove as much existing physiological noise as possible resulting in a clean ECG dataset. Then, four physiological noise types were applied with a specified signal-to-noise ratio (SNR). Hereafter, `datasets' refers to these six versions of the original ECG dataset: raw ECG signals, clean ECG signals and four applied ECG noise types. Each dataset contained signals for all 2678 subjects. Filtering of the ECG data was carried out using the ECGdeli toolbox for Matlab \cite{Pilia2020}. This filtering included: baseline wander removal, low pass filter (150 Hz), high pass filter (0.05 Hz), notch filter (49 Hz to 51 Hz) and isoline correction. 

The four physiological noise recordings were originally recorded and prepared for the MIT-BIH noise stress test database \cite{Moody1984,Goldberger2003} and were adapted for application to 12-lead ECG in a more recent study \cite{Petrenas2017}. Thirty minutes of each of the four noise types were available. Here, physiological noise was defined as noise that originated from the subject but excluded electrical activity from the heart. The four noise types were: baseline wander (bw), electrode movement (em), motion artefact (ma) and a linear combination of all three (all). Details of the six ECG datasets can be seen in Table \ref{tab:ECGdatasets}.

\begin{figure}[!htbp]
    \centering
    \includegraphics[width=.65\textwidth]{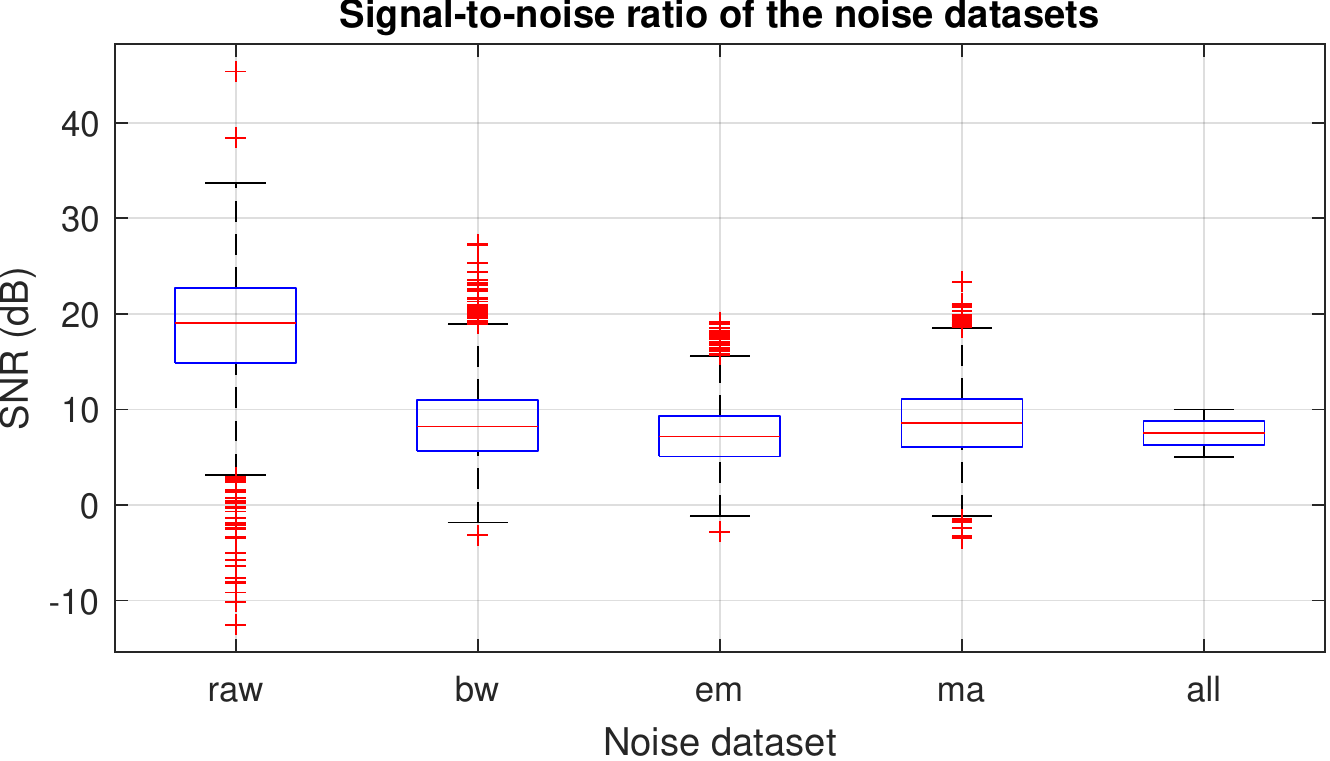}
    \caption{Signal-to-noise ratios (SNRs) for the noisy datasets. SNRs for the noise datasets calculated using the scaled noise signal and the clean signal. SNRs for the raw signal calculated using the clean signal and the difference between the raw and clean signals.}
    \label{fig:SNRs}
\end{figure}

Before the noise was added to the clean ECG dataset, it was scaled to give a signal-to-noise ratio (SNR) in a specified range. The raw ECG dataset had a high overall SNR (see Figure \ref{fig:SNRs}) and so was essentially free of noise, with the exception of a small number of records that contained a significant amount of noise. For the noisy datasets a SNR interval of 5 dB to 10 dB was specified to give a more uniform noise level and to ensure that the signals were sufficiently noisy for an effect to be detected. The procedure for applying noise to each of the 2678 clean signals was as follows: i) determine the length of the current clean ECG signal, ii) randomly select a segment of the same length from the thirty minutes of available noise, iii) randomly select a SNR between 5 and 10 dB, iv) using the clean signal, calculate the scaling factor required to scale the chosen `all' noise segment to the specified SNR, v) scale the `all' noise segment, and the three individual noise segments, using this scaling factor, vi) add the four noise segments to the clean signal to create four noisy signals. The resulting SNRs for the four noisy datasests can be seen in Figure \ref{fig:SNRs}. SNRs for the raw signal were calculated using the clean signal, and the difference between the raw and clean signal. 

\begin{table}[!htbp]
    \centering
    \caption{Details of the six ECG datasets including a Normal ECG signal example of each of the six variations. There are 2678 signals in each of the six datasets. Datasets are: raw, clean, baseline wander (bw), electrode movement (em), motion artefact (ma), combination of three noise types (all). Details of filtering are provided: isoline correction (Isoline corr.), baseline wander removal (Base. rem.), low pass filter (Lo pass), high pass filter (Hi pass), notch filter (Notch). Details of any added noise is provided. The clean ECG signal is shown in blue on all plots for comparison.}
    \label{tab:ECGdatasets}
    \begin{tabular}{>{\centering}m{0.8cm} m{4.5cm} m{2cm} m{3cm} }
        \textbf{Name} & \textbf{Example signal} & \textbf{Filtering} & \textbf{Added noise} \\
        \hline
        raw & \includegraphics[width=\linewidth]{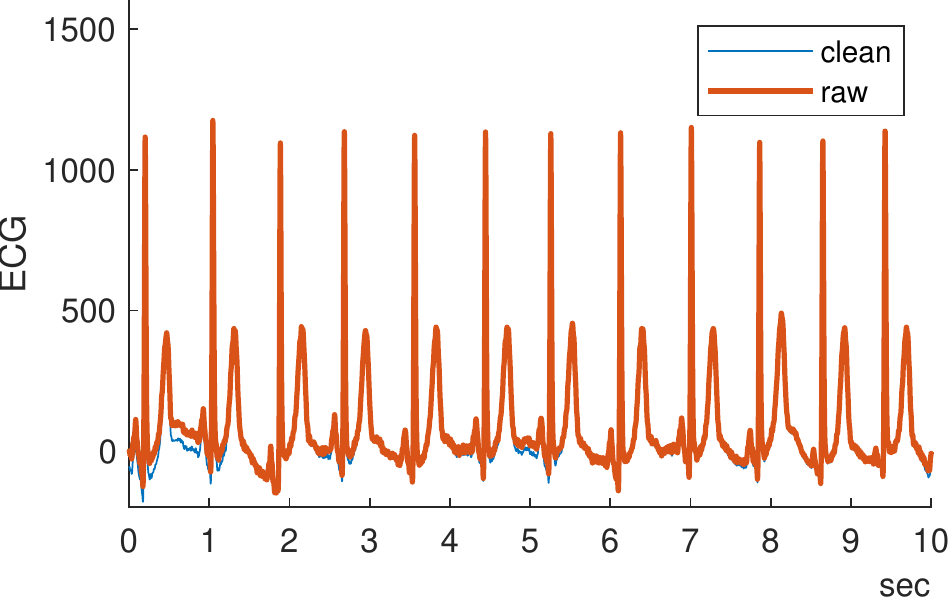} & Isoline corr. & None \\
        \hline
        clean & \includegraphics[width=\linewidth]{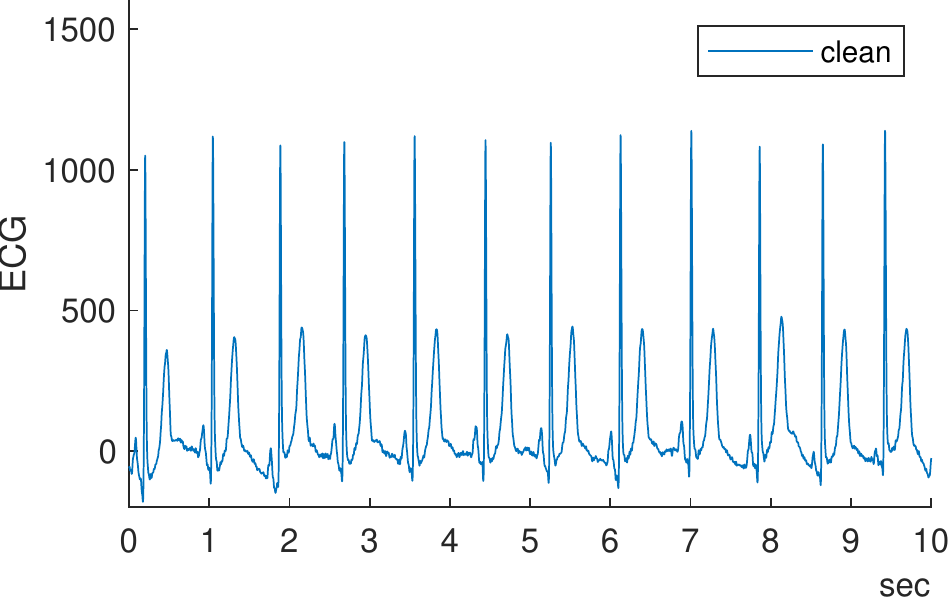} & Isoline corr.\newline Base. rem.\newline Lo pass\newline Hi pass\newline Notch & None \\
        \hline
        bw & \includegraphics[width=\linewidth]{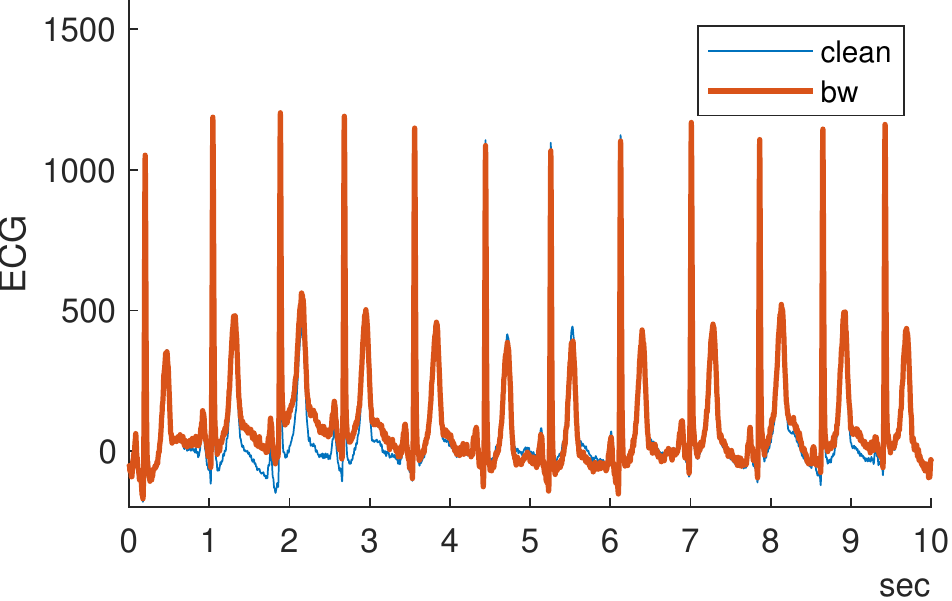} & Isoline corr.\newline Base. rem.\newline Lo pass\newline Hi pass\newline Notch & Baseline wander \\
        \hline
        em & \includegraphics[width=\linewidth]{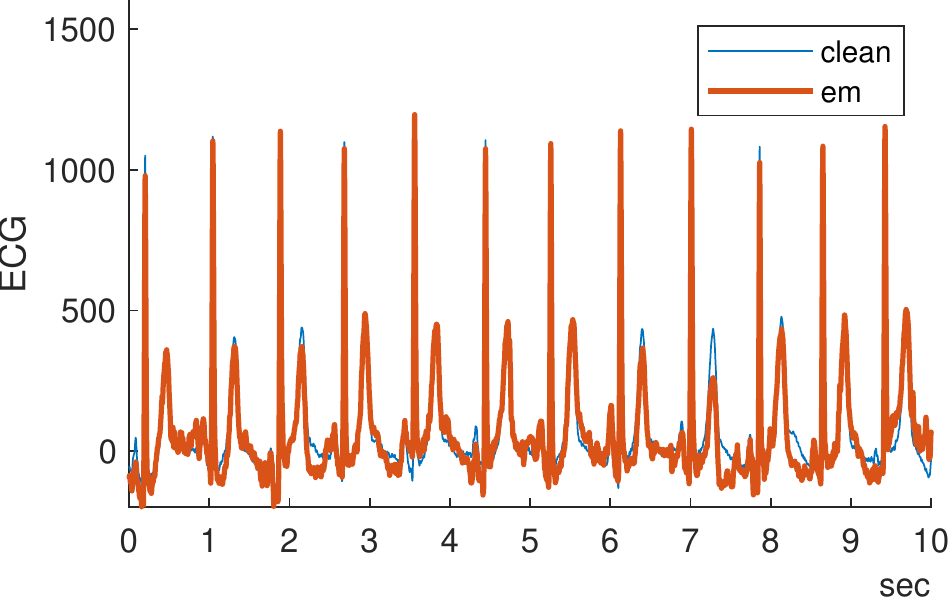} & Isoline corr.\newline Base. rem.\newline Lo pass\newline Hi pass\newline Notch & Electrode movement \\
        \hline
        ma & \includegraphics[width=\linewidth]{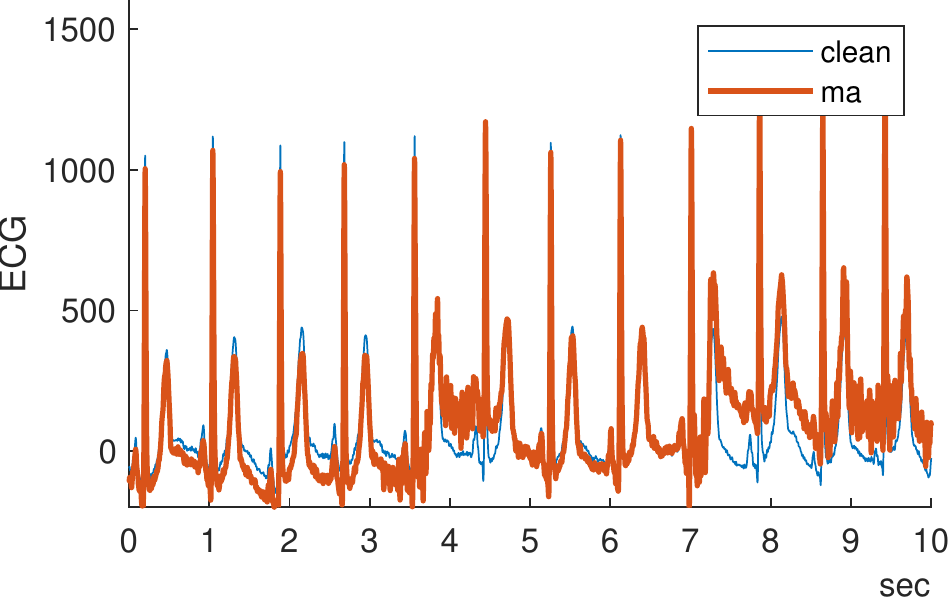} & Isoline corr.\newline Base. rem.\newline Lo pass\newline Hi pass\newline Notch & Motion artefact \\
        \hline
        all & \includegraphics[width=\linewidth]{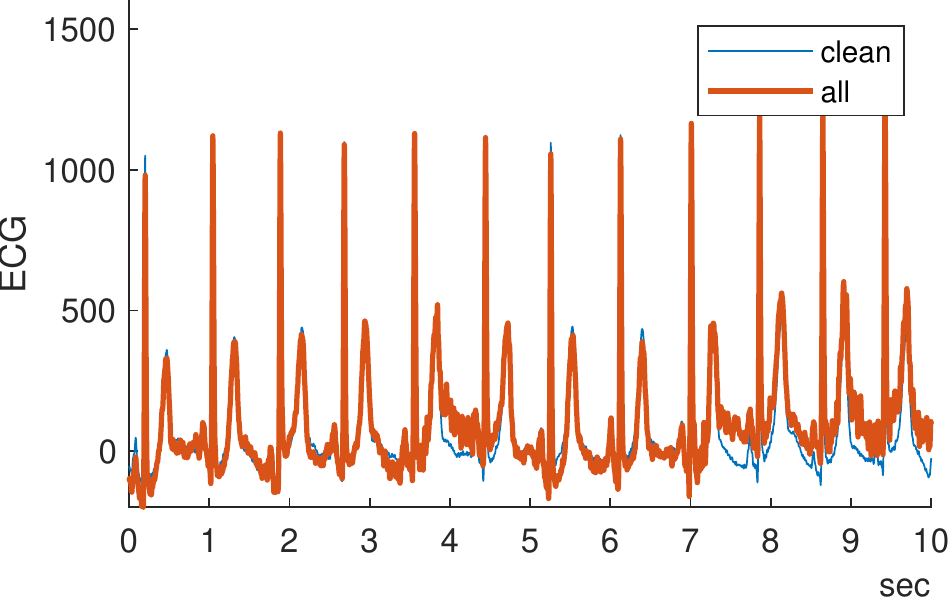} & Isoline corr.\newline Base. rem.\newline Lo pass\newline Hi pass\newline Notch & Baseline wander\newline Electrode movement\newline Motion artefact \\
        \hline
    \end{tabular}
\end{table}

\subsection{ECG image transforms}
Two transformation methods were used to generate ECG images for each of the six datasets: an attractor generated using the Symmetric Projection Attractor Reconstruction (SPAR) method \cite{Aston2018,Nandi2018} and a scalogram generated using the continuous wavelet transform. Lead II is commonly used to assess cardiac rhythm \cite{Meek2002} and was chosen to generate the ECG image transforms.

\begin{figure}[!htbp]
	\centering
	\begin{tabular}{>{\centering\arraybackslash}m{1.8cm} >{\centering\arraybackslash}m{4cm} >{\centering\arraybackslash}m{2cm} >{\centering\arraybackslash}m{2cm}}
	\textbf{Class} & \textbf{ECG signal} & \textbf{Attractor} & \textbf{Scalogram} \\
	Normal & \includegraphics[height=2cm]{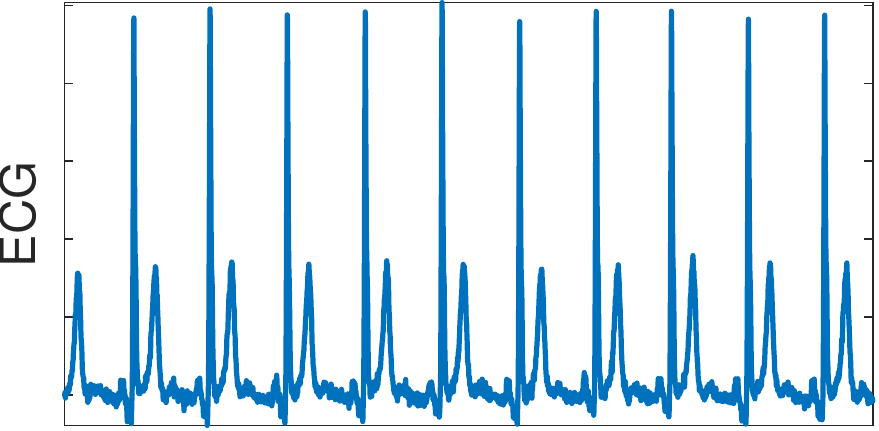} & \includegraphics[height=2cm]{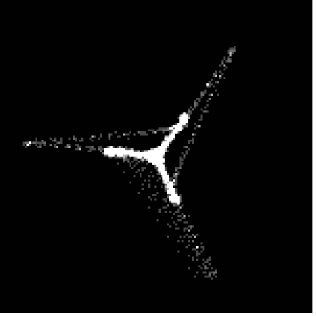} & \includegraphics[height=2cm]{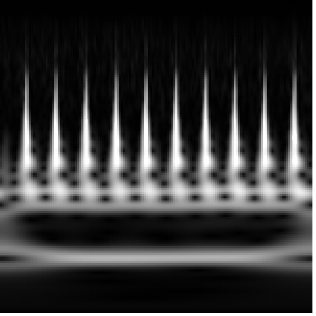} \\
	AF & \includegraphics[height=2cm]{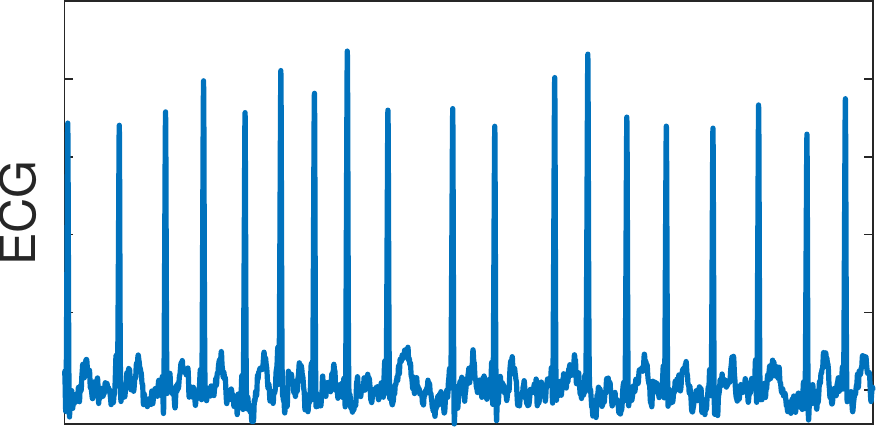} & \includegraphics[height=2cm]{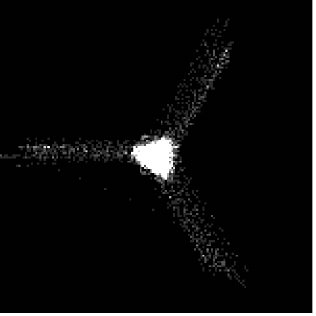} & \includegraphics[height=2cm]{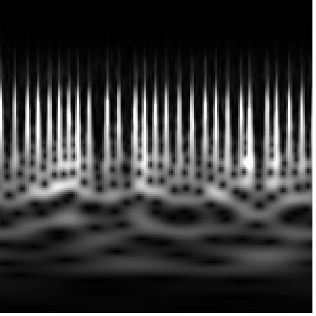} \\
	STD & \includegraphics[height=2cm]{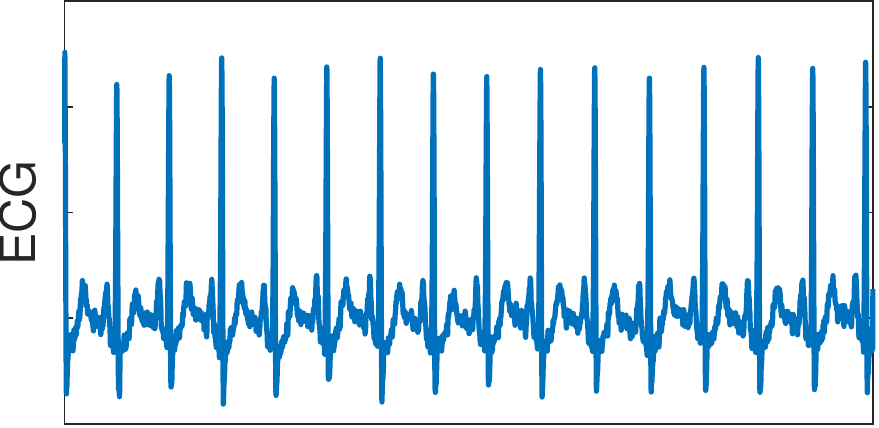} & \includegraphics[height=2cm]{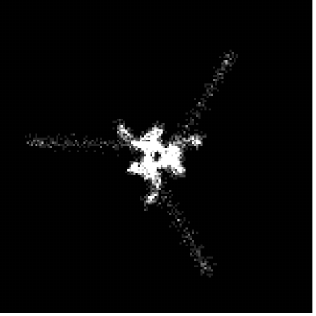} & \includegraphics[height=2cm]{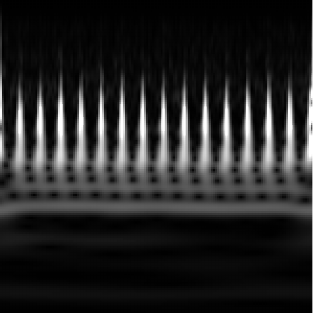} \\
	\end{tabular}
	\caption{Representative ECG signal for each class, along with the corresponding attractor and scalogram. Note that attractor images have increased constrast for better visibility.}\label{fig:ExampleImages}
\end{figure}

The SPAR method uses delay coordinates to replot the data in three dimensions, then projects the resulting three dimensional plot to two dimensions from which a density plot is generated. The resulting shape is known as an attractor. An ECG attractor captures morphology information of the signal but factors out heart rate. The continuous wavelet transform was generated using an analytic Morse wavelet, with 16 wavelet bandpass filters per octave and the scalogram was obtained by plotting the absolute value of the resulting coefficients. An ECG scalogram captures frequency domain information of the signal. See Figure \ref{fig:ExampleImages} for a representative ECG, attractor and scalogram for each of the three classes. Both image types were generated with dimension \(150 \times 150\). All image transforms were grayscale and were generated using the full ECG signal. As two image transformations were applied to each of the six datasets, there were twelve image datasets in total. Hereafter `image dataset' refers to one of these twelve.

\subsection{Transfer learning with CNNs}
To classify the ECG image transforms, four CNNs pretrained using ImageNet were initially assessed: AlexNet, GoogLeNet, VGG-16 and ResNet-50. Results using these networks identified ResNet-50 as the best performing and so this network was used for the main study. Transfer learning was used to adapt each of the networks for the image transforms. To do this all layers of each network were frozen and the last learnable layer and classification layer were replaced and retrained using the ECG image transforms. All image transforms were rescaled to either \(224\times224\) or \(227\times227\) depending on the requirements of the network input layer.

\begin{figure}[!htbp]
    \centering
    \includegraphics[width=.7\linewidth]{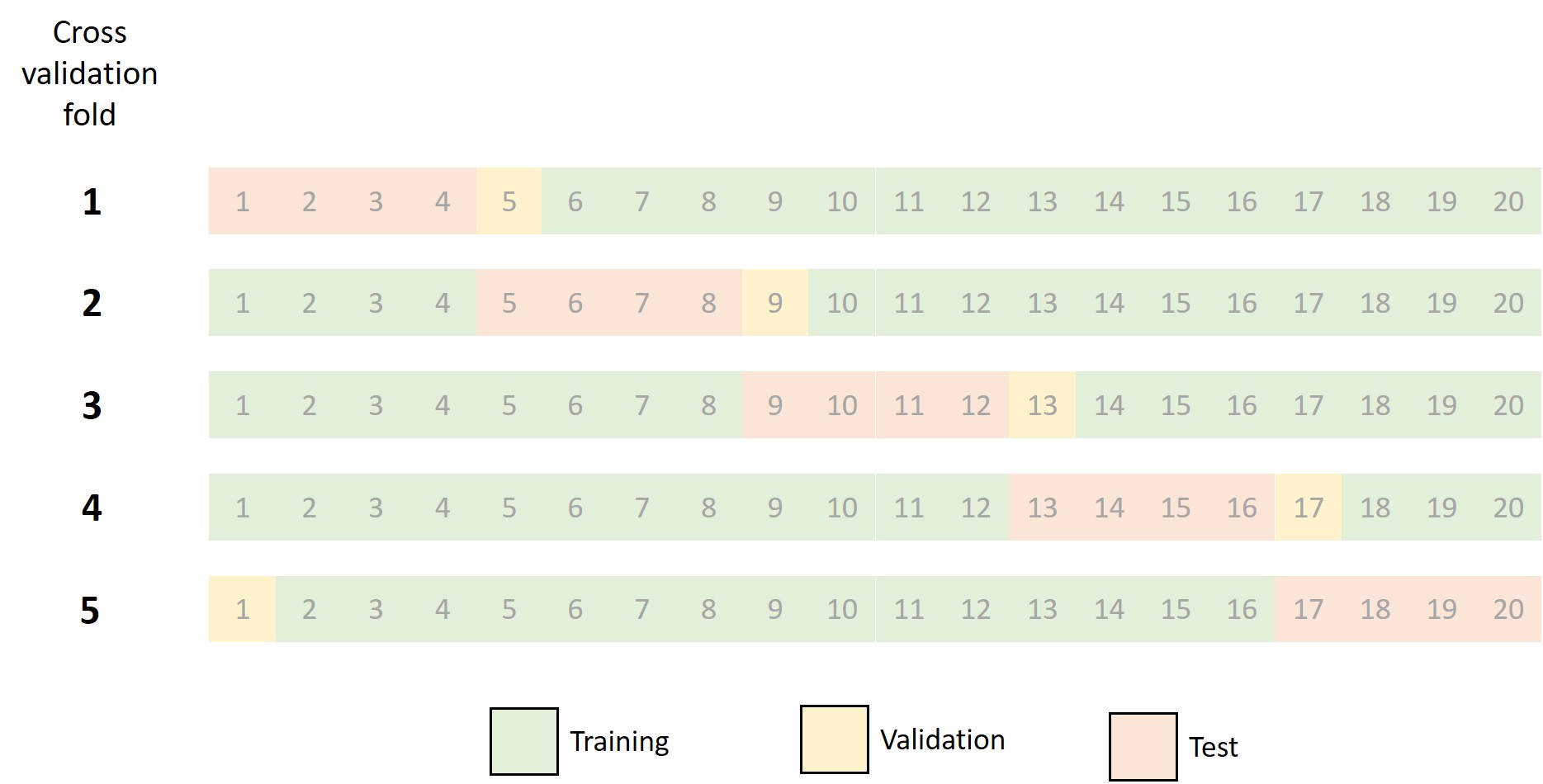}
    \caption{Diagram of how the data was split into training (75\%), validation (5\%) and test (20\%) data for each of the five cross validation folds.}
    \label{fig:CrossValidation}
\end{figure}

Each of the twelve image datasets were used to train the network. Five fold cross validation was used. For each cross validation fold, the data was split into training (75\%), validation (5\%) and test (20\%) data (Figure \ref{fig:CrossValidation}). Predefined cross validation folds were used, thus giving consistency for the different datasets.  To prevent overfitting, networks were set to stop training when the validation loss was larger than the previous smallest value five times consecutively.

To carry out the study of robustness, and investigate the impact that physiological ECG noise has on classification performance, three variations of the network training and testing were carried out:
\begin{enumerate}
    \item Network training and validation using a single image dataset, used to classify test data from the same image dataset;
    \item Network training and validation using a clean ECG image dataset, used to classify test data from all other image datasets;
    \item Network training and validation using a noisy ECG image dataset, used to classify test data from all other image datasets.
\end{enumerate}

\subsection{Performance metrics}
As the classes were slightly imbalanced, the F1 score was used to evaluate the performance of the networks. The F1 score is the harmonic mean of precision and recall and is calculated as follows:
\[
F_1 = 2\cdot \frac{\textrm{precision}\cdot \textrm{recall}}{\textrm{precision} + \textrm{recall}}
\]
where precision is the fraction of predicted labels of a specific class that are correct and recall is the fraction of a specific class that is correctly identified. To account for the multiple classes, the macro averaged F1 score was used, whereby the F1 score was calculated for each class individually and the arithmetic mean of these was taken. Unless specified, the F1 scores reported below refer to overall network performance (macro averaged F1 score) as opposed to individual class F1 score.

\section{Results}\label{sec:Res}

\subsection{Network comparison: Training and test data from same dataset}\label{sec:NetComp}
\begin{figure}[!h] 
	\centering
	\begin{subfigure}[t]{.5\linewidth}
		\centering
		\includegraphics[width=\textwidth]{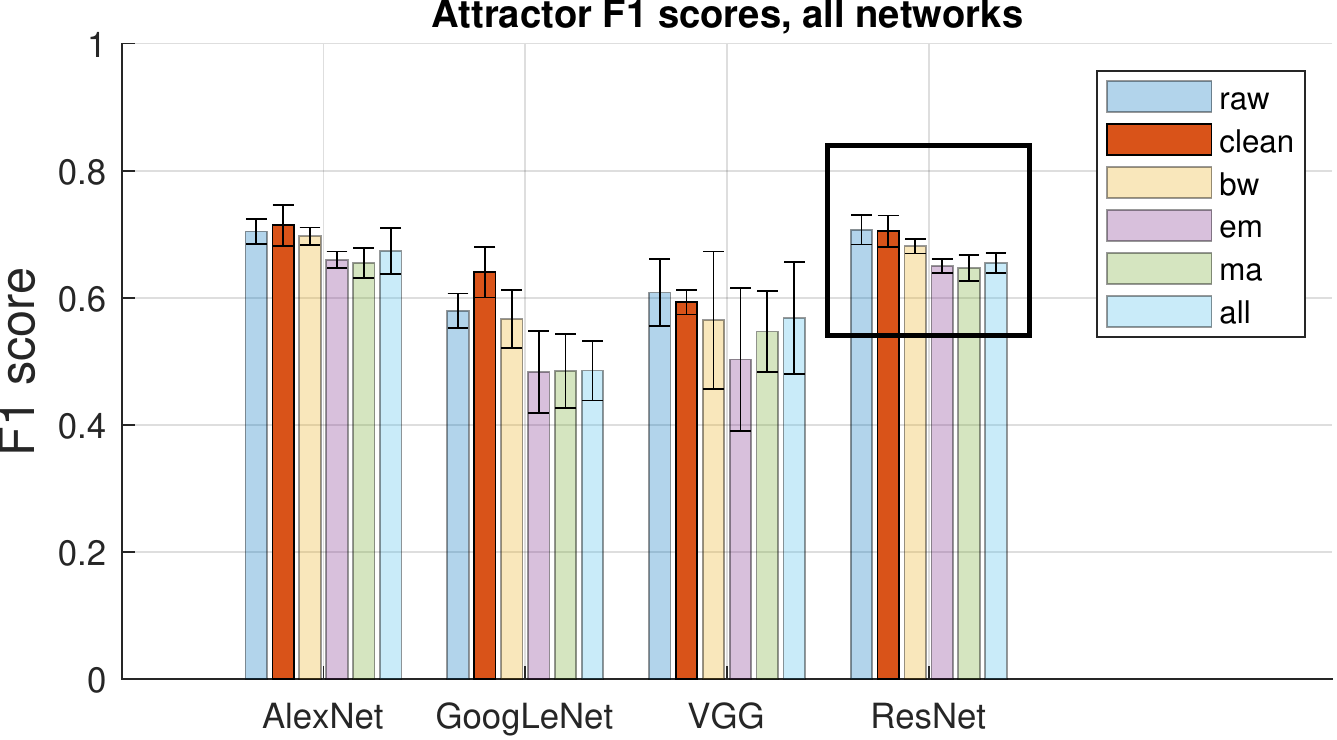}
		\caption{}\label{fig:AttNetComp}
	\end{subfigure}%
	\begin{subfigure}[t]{.5\linewidth}
		\centering
		\includegraphics[width=\textwidth]{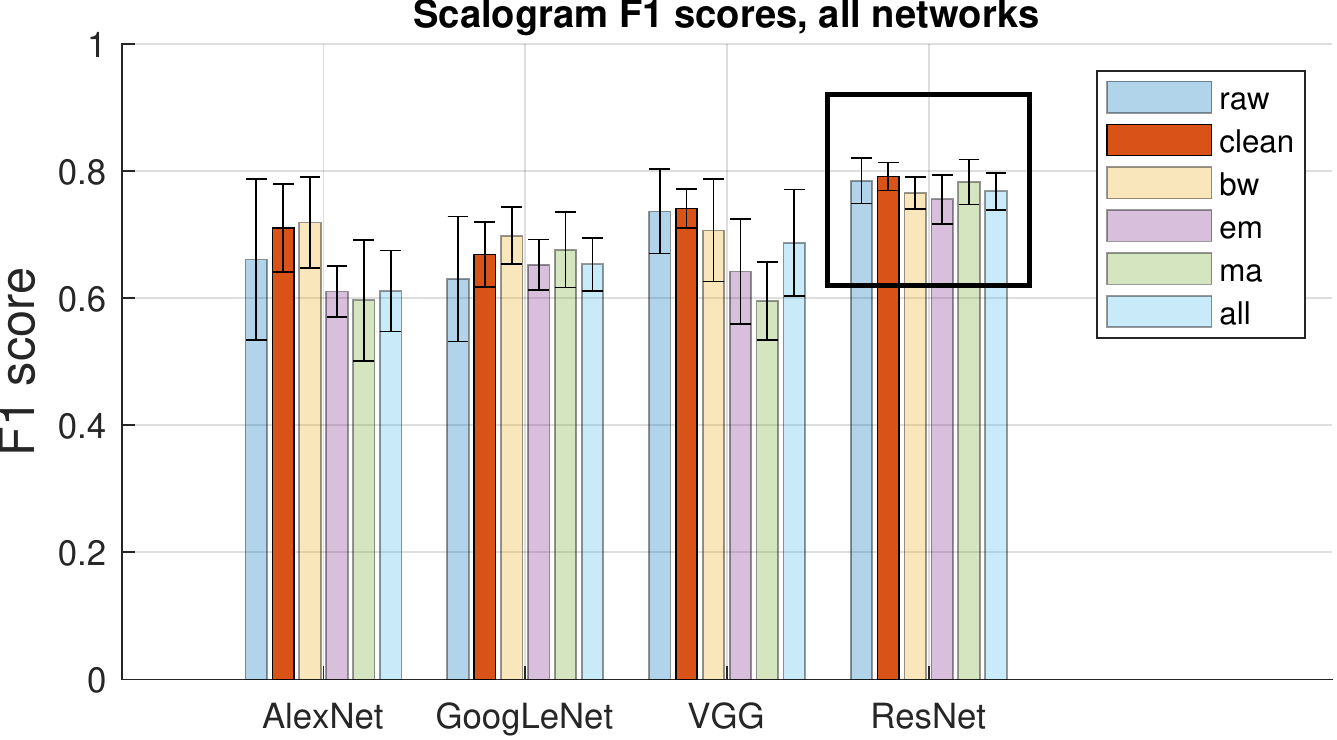}
		\caption{}\label{fig:SclNetComp}
	\end{subfigure}
	\caption{Comparison of network performance when training and test data were taken from the same dataset, for \subref{fig:AttNetComp}) the attractor images and \subref{fig:SclNetComp}) the scalogram images. Each cluster of bars shows F1 scores for a single network, where each bar represents an ECG dataset: raw, clean, baseline wander (bw), electrode movement (em), motion artefact (ma) and all three combined (all). The raw data performance (raw) is shown for comparison. Error bars show standard deviation across the five cross validation folds. Black boxes highlight results for the chosen network.}\label{fig:NetComp}
\end{figure}

Overall network performance for four different networks trained and tested on each of the six datasets can be seen in Figure \ref{fig:NetComp}. Using F1 scores as a metric of overall network performance, ResNet-50 is the best performing network and has the lowest standard deviation across the five cross validation folds, although for the attractor images AlexNet has similar performance. Furthermore, unlike the other networks, ResNet-50 performance is fairly consistent across the datasets, with all F1 scores in the range 0.65 to 0.71 for the SPAR images, and in the range 0.76 to 0.79 for the scalogram images. ResNet-50 has been used for the remainder of the analysis.

\subsection{Noise comparison: Training and test data from the same dataset}
When networks were trained and tested using data from the same dataset, classification performance F1 scores for the noisy datasets were consistently lower than for the clean dataset. These results can be seen within the black boxes in Figure \ref{fig:NetComp}. It should be noted that the error bars overlap more for scalogram than for the attractor images, suggesting that the performance decrease with noise is less significant. F1 score changes from the clean dataset to the all noise dataset were 0.70 to 0.65 for the attractor images and 0.79 to 0.77 for the scalogram images. ResNet-50 appeared to perform most consistently across all noise types, with the smallest F1 score decrease in performance. Other networks showed larger changes and in some cases performance increased in the presence of noise (see Figure \ref{fig:SclNetComp}, AlexNet for example). 

\subsection{Using a network trained on clean data to classify noisy data}\label{sec:cleanclassifynoisy}
\begin{figure}[!h] 
	\centering
	\includegraphics[width=.6\linewidth]{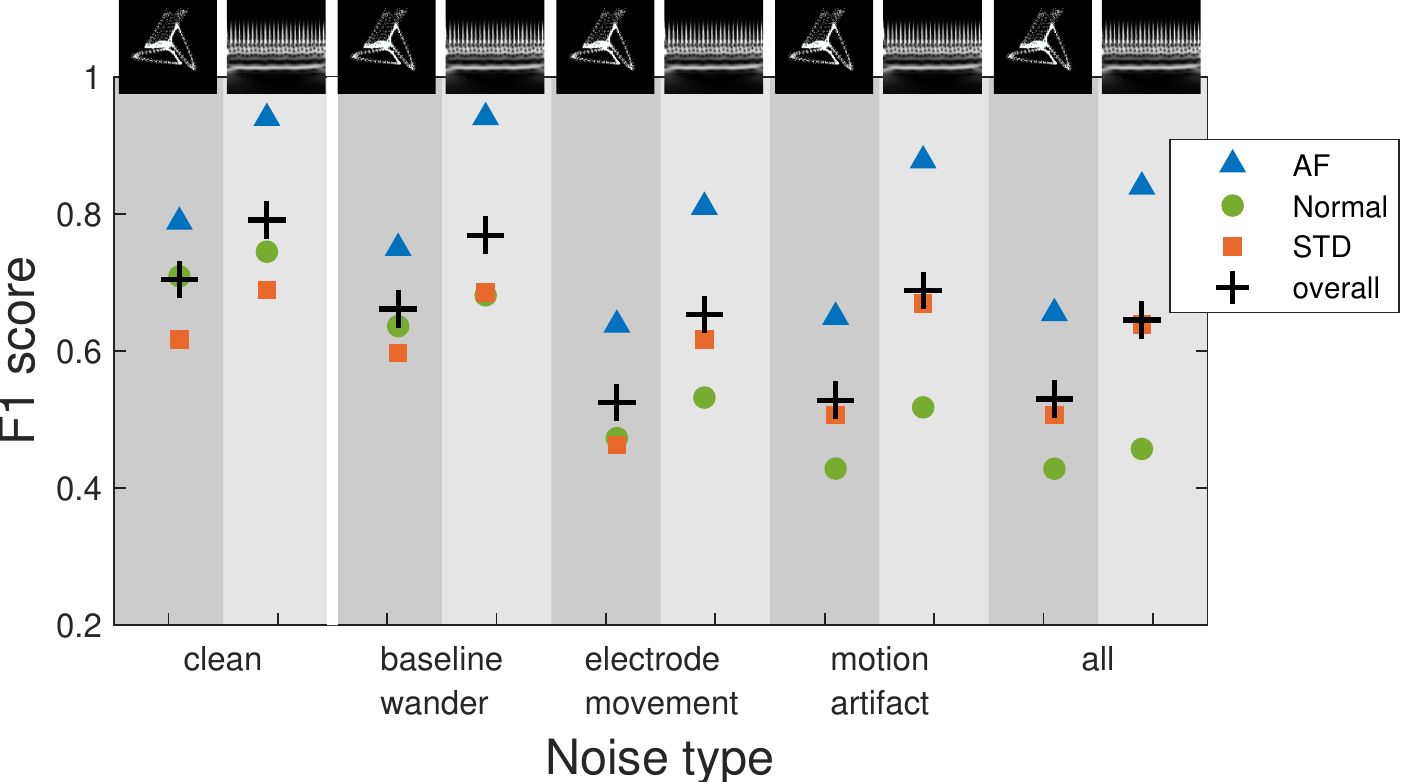}
		\caption{Breakdown of ResNet-50 F1 scores by class for the network trained using the clean dataset then tested using the noisy datasets. Here \(+\) is overall network F1, \(\triangle\) is AF, {\Large\(\circ\)} is Normal and \(\square\) is STD. Attractor and scalogram F1 scores are shown side by side, indicated by the background shading. From left to right, results are shown for clean, baseline wander, electrode movement and all three combined.}\label{fig:ClassMetricsCleanNoisy}
\end{figure}

ResNet-50 was chosen for further study. When the network that had been trained on the clean ECG image dataset was used to classify test data from the noisy datasets, F1 scores decreased by up to \(0.2\). A summary of the overall F1 scores is given in Table \ref{tab:F1summaryCleanNoisy}. A summary of results, including a breakdown of the F1 scores by class, can be seen in Figure \ref{fig:ClassMetricsCleanNoisy}.

\begin{table}[!h] 
	\caption{Summary of overall network F1 scores for the ResNet-50 network trained using the clean dataset, then tested using the noisy datasets. Values shown are the arithmetic mean and standard deviation across the five cross validation folds. Results shown for attractor (att) and scalogram (scl) image transforms. Note that results for clean data are a duplication of data shown in Table \ref{tab:F1summary}. The results for raw data are shown for comparison.}\label{tab:F1summaryCleanNoisy}
	
	\centering
	\begin{tabular}[t]{l c || c c c c c}
		\multicolumn{6}{c}{Overall} \\
		\hline
		\textbf{image} & \textbf{clean} & \textbf{\textit{raw}} & \textbf{bw} & \textbf{em} & \textbf{ma} & \textbf{all} \\
		\hline
		\textbf{att} & 0.70 (0.02) & 0.69 (0.03) & 0.66 (0.04) & 0.52 (0.05) & 0.53 (0.05) & 0.53 (0.05) \\
		\textbf{scl} & 0.79 (0.02) & 0.78 (0.02) & 0.77 (0.02) & 0.65 (0.02) & 0.69 (0.02) & 0.65 (0.01) \\
		\hline
	\end{tabular}%

\end{table}

\subsection{Using a network trained on noisy data to classify clean and noisy data}\label{sec:noisyclassifyclean}
\begin{figure}[!h] 
	\centering
	\includegraphics[width=.6\linewidth]{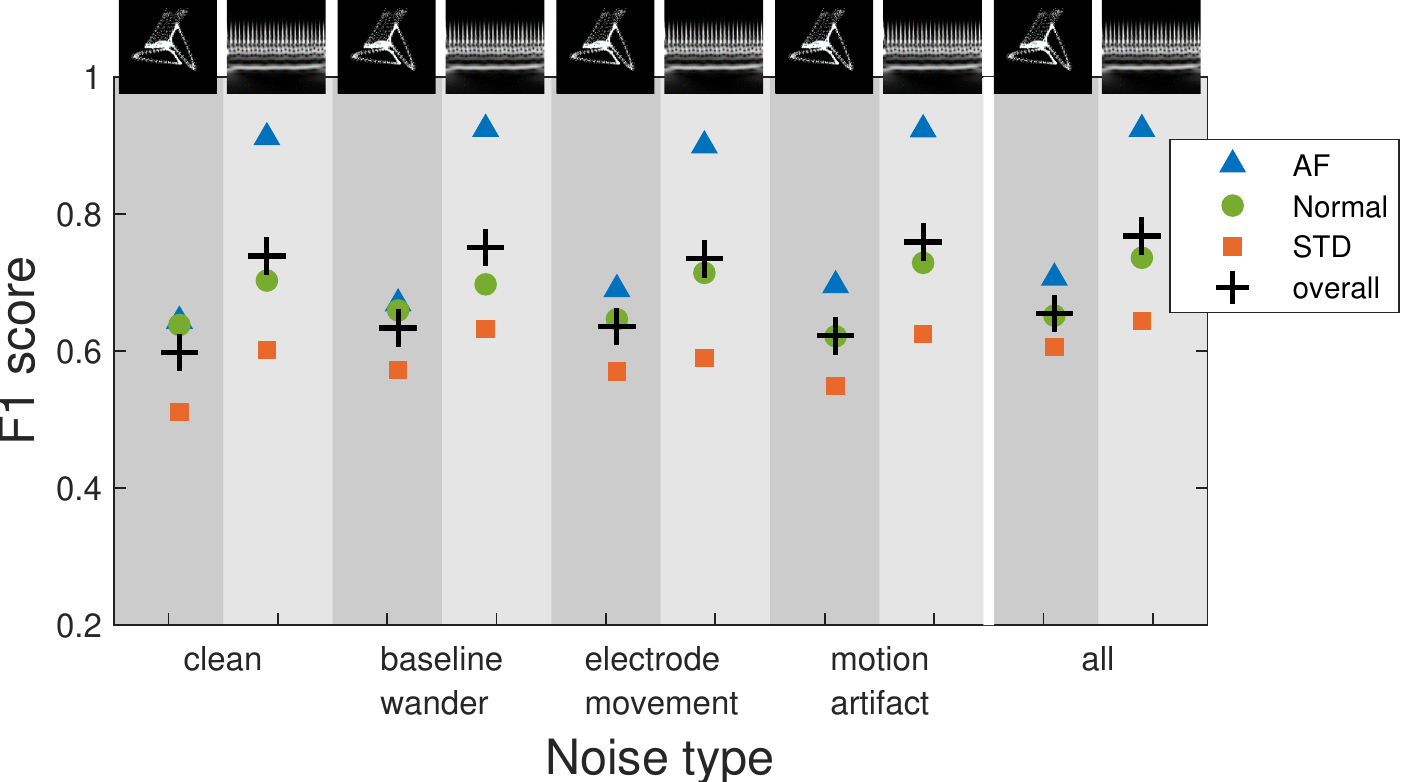}
	\caption{Breakdown of ResNet-50 F1 scores by class for the network trained using the all noise dataset then tested using all other datasets. Here \(+\) is overall network F1, \(\triangle\) is AF, {\Large\(\circ\)} is Normal and \(\square\) is STD. Attractor and scalogram F1 scores are shown side by side, indicated by the background shading. From left to right, results are shown for clean, baseline wander, electrode movement and all three combined.}\label{fig:ClassMetricsNoisyClean}
\end{figure}

A network that had been trained on the all noise ECG image dataset was used to classify test data from the clean and noisy datasets. See the overall F1 scores in Table \ref{tab:F1summaryNoisyClean} and a summary of results for all classes in Figure \ref{fig:ClassMetricsNoisyClean}. In this case, the performance decrease was smaller than when the network trained on clean data was used to classify noisy data. The all noise ECG image dataset performed fairly consistently in classifying all other ECG image datasets.

\begin{table}[!h] 
	\caption{Summary of overall network F1 scores for the ResNet-50 network trained using the all noise image dataset then tested using all other datasets. Values shown are the arithmetic mean and standard deviation across the five cross validation folds. Results shown for attractor (att) and scalogram (scl) images. Note that results for all noise are a duplication of data shown in Table \ref{tab:F1summary}. The results for raw data are shown for comparison.}\label{tab:F1summaryNoisyClean}
	\centering
	\begin{tabular}[t]{l c c c c c || c}
		\multicolumn{7}{c}{Overall} \\
		\hline
		\textbf{image} & \textbf{\textit{raw}} & \textbf{clean} & \textbf{bw} & \textbf{em} & \textbf{ma} & \textbf{all} \\
		\hline
		\textbf{att} & 0.60 (0.05) & 0.60 (0.04) & 0.63 (0.04) & 0.64 (0.02) & 0.62 (0.03) & 0.65 (0.02) \\
		\textbf{scl} & 0.74 (0.03) & 0.74 (0.03) & 0.75 (0.02) & 0.73 (0.02) & 0.76 (0.03) & 0.77 (0.03) \\
		\hline
	\end{tabular}%
	
\end{table}

\subsection{Class comparison: Training and test data from the same dataset}\label{sec:ClassPerf}
\begin{figure}[!h]
	\centering
	\includegraphics[width=.6\linewidth]{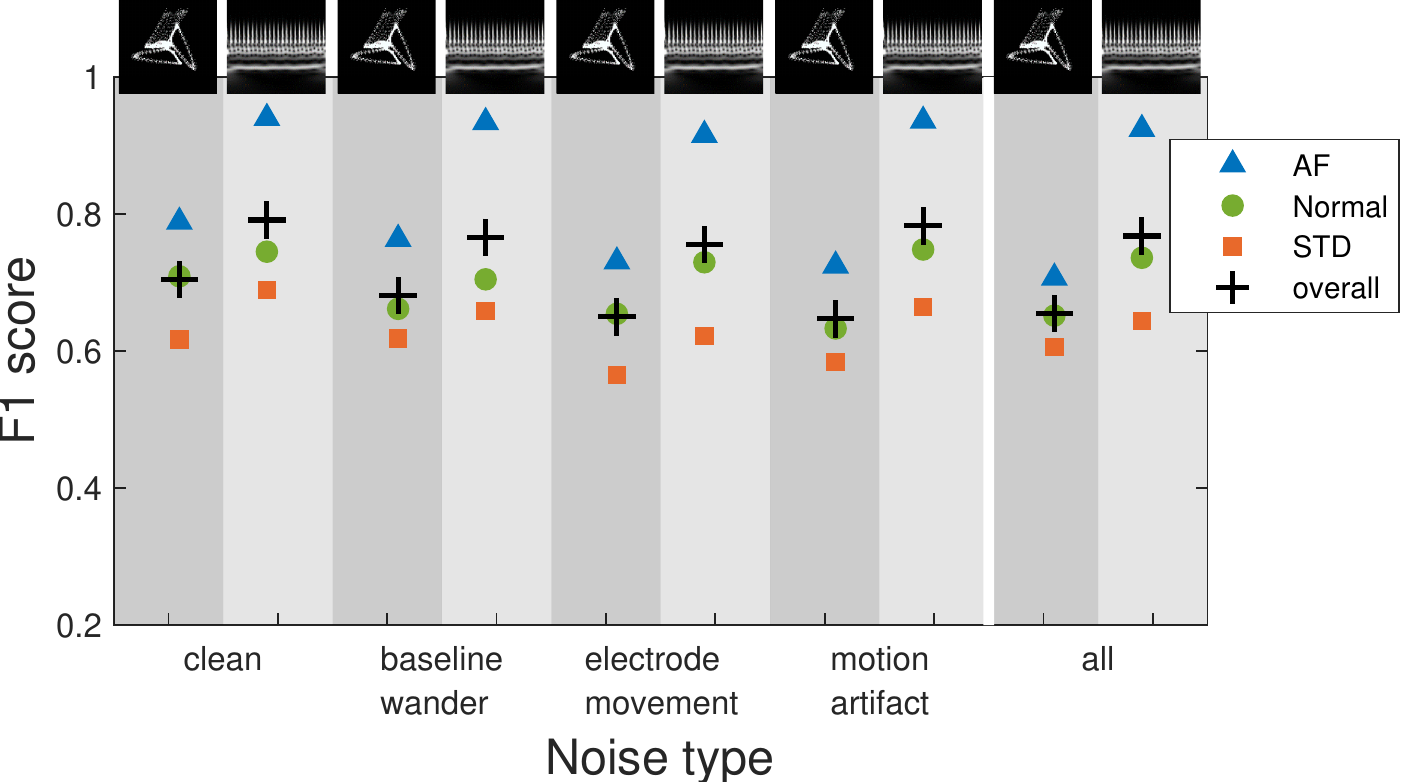}
	\caption{Breakdown of ResNet-50 F1 scores by class when training and test data was taken from the same dataset. Here \(+\) is the overall network F1 score, \(\triangle\) is AF, {\Large\(\circ\)} is Normal and \(\square\) is STD. Attractor and scalogram F1 scores are shown side by side, indicated by the background shading. Results for clean data are shown in the left hand columns, followed by results for baseline wander, electrode movement, motion artefact and all three combined.}\label{fig:ClassMetrics}
\end{figure}

The F1 scores seen so far have been the macro averaged (overall network) F1 scores, that is the arithmetic mean of the F1 scores for each of the three classes individually. Individual class F1 scores for networks whose training and test data were taken from the same ECG dataset can be seen in Figure \ref{fig:ClassMetrics}. A summary of F1 scores for each class individually can be seen in Table \ref{tab:F1summary}.

\begin{table}[!h] 
	\caption{Summary of overall network F1 scores for the ResNet-50 network whose training and test data was taken from the same dataset. Values shown are the arithmetic mean and standard deviation across the five cross validation folds. Results shown for attractor (att) and scalogram (scl) images. From top to bottom the table shows results for the overall network performance on all three classes, for AF (atrial fibrillation) only, Normal only and STD (ST depression) only. Results with the raw dataset are included for comparison.}\label{tab:F1summary}
	\centering
	\begin{tabular}[t]{l c c c c c c}
		\multicolumn{7}{c}{Overall} \\
		\hline
		\textbf{image} & \textbf{\textit{raw}} & \textbf{clean} & \textbf{bw} & \textbf{em} & \textbf{ma} & \textbf{all} \\
		\hline
		\textbf{att} & 0.71 (0.02) & 0.70 (0.02) & 0.68 (0.01) & 0.65 (0.01) & 0.65 (0.02) & 0.65 (0.02) \\
		\textbf{scl} & 0.78 (0.04) & 0.79 (0.02) & 0.77 (0.03) & 0.76 (0.04) & 0.78 (0.04) & 0.77 (0.03) \\
		\hline
	\end{tabular}%
	\vspace{1em}
	
	{\scriptsize
		\begin{tabular}[t]{l c c c c c c}
			\multicolumn{7}{c}{AF} \\
			\hline
			\textbf{image} & \textbf{\textit{raw}} & \textbf{clean} & \textbf{bw} & \textbf{em} & \textbf{ma} & \textbf{all} \\
			\hline
			\textbf{att} & 0.78 (0.01) & 0.79 (0.04) & 0.76 (0.02) & 0.73 (0.03) & 0.72 (0.03) & 0.71 (0.04) \\
			\textbf{scl} & 0.94 (0.01) & 0.94 (0.02) & 0.93 (0.02) & 0.92 (0.02) & 0.94 (0.02) & 0.92 (0.02) \\
			\hline
		\end{tabular}
		
		\vspace{1em}
		\begin{tabular}[t]{l c c c c c c}
			\multicolumn{7}{c}{Normal} \\
			\hline
			\textbf{image} & \textbf{\textit{raw}} & \textbf{clean} & \textbf{bw} & \textbf{em} & \textbf{ma} & \textbf{all} \\
			\hline
			\textbf{att} & 0.70 (0.04) & 0.71 (0.03) & 0.66 (0.04) & 0.65 (0.03) & 0.63 (0.05) & 0.65 (0.04) \\
			\textbf{scl} & 0.75 (0.03) & 0.75 (0.03) & 0.70 (0.05) & 0.73 (0.03) & 0.75 (0.05) & 0.74 (0.03) \\
			\hline
		\end{tabular}
		
		\vspace{1em}
		\begin{tabular}[t]{l c c c c c c}
			\multicolumn{7}{c}{STD} \\
			\hline
			\textbf{image} & \textbf{\textit{raw}} & \textbf{clean} & \textbf{bw} & \textbf{em} & \textbf{ma} & \textbf{all} \\
			\hline
			\textbf{att} & 0.64 (0.02) & 0.62 (0.04) & 0.62 (0.03) & 0.57 (0.03) & 0.58 (0.04) & 0.61 (0.02) \\
			\textbf{scl} & 0.66 (0.07) & 0.69 (0.03) & 0.66 (0.05) & 0.62 (0.07) & 0.66 (0.04) & 0.64 (0.06) \\
			\hline
		\end{tabular}
	}
\end{table}

\section{Discussion}
Deep learning is a well established and rapidly growing approach for ECG classification \cite{Hong2020} where classification performance is often determined by both the quality of the training data used and the chosen model architecture. In this paper, a ResNet-50 architecture has been used to classify SPAR attractor and scalogram image transforms of an ECG signal. We have explored how robust the networks are to physiological ECG noise, and also how including or excluding physiological noise from the training data impacts robustness. The key findings with regards to choice of training data can be summarised as follows:
\begin{enumerate}
    \item \textbf{Train network using clean data if filtering of future unseen (test) data is known.} A network trained and tested on one data type (clean or noisy) has better performance than a network trained on clean data and tested on noisy data or vice versa. Therefore, for example, if it is known that all data the network will have to classify in the future will be filtered, then use filtered data to train the network.
    \item \textbf{Train network using noisy data if filtering of future unseen (test) data is unknown.} A network trained on noisy data and tested on clean data has a smaller performance decrease than a network trained on clean data and tested on noisy data.
\end{enumerate}

It is perhaps unsurprising that adding physiological ECG noise to the data on which the network is trained makes the network more robust and generalisable than a network trained on clean data alone, as adding (non-physiological) noise to training data is a known method for improving deep network robustness and preventing overfitting \cite{Goodfellow2016,An1996}.

\subsection{Robustness}
Noise and interference are common causes of ECG misclassification \cite{Luo2010} and it is beneficial to gain an understanding of how physiological ECG noise affects deep network robustness. Here we have demonstrated to what degree ECG classification performance was decreased in the presence of different physiological noise types, and found that electrode movement and motion artefact have a bigger impact on performance than baseline wander for attractor images, while networks trained using scalogram image transforms appear to be more robust (see Table \ref{tab:F1summary}, overall network scores). The SPAR attractor method inherently factors out baseline wander (see \cite{Nandi2018} for details) which may account for the smaller performance decrease for baseline wander. This result for baseline wander may also be partially due to the fact that some of the ECG signals were short (\(<\)10 s) so only a minimal amount of baseline wander could be applied.

When the network trained on clean data was used to classify images from the different noisy datasets, performance measured by F1 score decreased minimally for both image types when baseline wander was applied (\(<0.04\)). When motion artefact, electrode movement or all types of noise combined were added to the input data, classification performance using the network trained on clean data decreased by \(\sim0.1\) for the scalograms and \(\sim0.2\) for the attractors. A possible explanation for the higher robustness of scalogram image transforms to the motion artefact and electrode movement noise could be that the relevant frequency information for these noise types is distinct from the frequency information used for classification in these images. We also note that performance for the raw data was very similar to that for the clean data in all cases, which is not surprising as most signals contained very little noise (Figure \ref{fig:SNRs}).

Eventually, ECG classification networks may be used in a clinical setting and the type or level of noise in this setting will be unknown beforehand. In this situation, it is beneficial to have prior understanding of the robustness of a trained network. While the best performance was found using a network trained on data with the same type of noise as the input data, it is noteworthy that the `penalty' of using a network trained on noisy data was much less than when using a network trained on clean data to classify noisy data (Sections \ref{sec:cleanclassifynoisy} and \ref{sec:noisyclassifyclean}). Interestingly, the classification of Normal ECG signals was most affected when using the network trained on clean data to classify noisy data (Figure \ref{fig:ClassMetricsCleanNoisy}). Overall the network appears to be more robust to unseen noise on the input data when noise has been included in the training data.

While these results reveal interesting effects of noise on classification, Figure \ref{fig:NetComp} demonstrates the strong dependence on the choice of network and that further work is required in this area. Performance differences in the presence of noise for different networks suggest the need to test individual network plus dataset combinations for robustness and, perhaps more importantly, to develop datasets and techniques for evaluating robustness of networks. Work is underway in this area (see \cite{Strodthoff2020} and \cite{Han2020}, for example).

Resistance to adversarial attacks is an increasingly desirable quality as deep learning models for ECG classification become more widely utilised, especially in a clinical setting. Imperceptible (to a human) changes in an ECG have been shown to cause misclassification by a deep network \cite{Han2020}, but would likely have little effect on the image transforms. Adversarial training, where intentionally placed noise is used to generate additional input data, has been shown to increase network stability \cite{Kurakin2017}. The use of ECG image transforms (rather than the raw signal) in deep learning models may be one way of increasing robustness of a network to adversarial attacks, but further work is required in this area to quantify how robust image transformation methods are to imperceptible changes on the original ECG, compared to methods using the original ECG.

\subsection{ECG preprocessing}
Prior to image transformation, the ECG signals were filtered to remove as much real physiological noise as possible before applying the noise types. As real ECG signals were used, the amount of original noise remaining in each clean signal before the noise was applied was different despite the same method being used to filter all ECG signals. One implication of this situation is that, while we can use results of this study to draw conclusions about the impact of various noise types individually, it should be remembered that there may be traces of other noise types in the underlying ECG signal in addition to the newly applied physiological noise. This result echoes the findings of the original physiological noise dataset, where it was noted that it was almost impossible to record electrode movement or motion artefact without baseline wander \cite{Moody1984}, suggesting that the results on electrode movement or motion artefact alone are less meaningful.

An alternative approach to the generation of the noisy datasets would be to use clean synthetic ECG signals and add the noise to these. This approach would ensure as far as possible that there is only one type of noise being applied to the ECG when that is intended. However, a synthetic ECG dataset may not be as representative of real ECG data and it may be more difficult to translate findings using a synthetic dataset than real dataset. Further work is required in this area. Finally, other studies have shown a decrease in performance with increasing SNR \cite{Oster2015}. Here, the physiological noise was applied with a small, fixed range for the SNR and further work is required to understand how different magnitude SNR noise affects performance. In particular, with regards to our finding that including noise in the training data increases network robustness it would be valuable to determine levels of noise required.

\subsection{Image transforms}
A notable difference in performance between the frequency based scalogram image transforms and the morphology based SPAR images was for atrial fibrillation (AF), see Figure \ref{fig:ClassMetrics}, where the scalogram F1 score for AF was at least \(0.15\) higher than the attractor F1 score for all noise types. AF is characterised by irregular \textit{f} waves, known as fibrillatory waves \cite{Goldberger2018}, which are expected to be more easily identifiable on the scalogram, suggesting that machine learning methods incorporating frequency based techniques are better suited to identifying conditions characterised by frequency based features such as fibrillatory waves. Amongst the criteria for diagnosis of AF is a heart rate above 100 bpm \cite{Goldberger2018}, and this information would be captured in the scalogram image but not in the attractor image, which factors out heart rate and focusses exclusively on the ECG waveform morphology. Future models that combine the attractor image with additional information such as heart rate may improve the performance of the attractor for AF.

In contrast, classification performance for STD using attractor images is similar to that for scalogram images, despite having a lower overall performance. STD is characterised by an ST segment below the baseline of the ECG which affects the morphology of the ECG waveform. The attractor captures ECG waveform morphology in detail \cite{Nandi2018}, which may explain why ST segment depression is not too difficult to identify using SPAR images.

Signals in the dataset were between 8 and 138 seconds long and this may have impacted classification performance. The SPAR image transformation method normalises for signal length, whereas scalogram images for signals with different lengths appear different and can potentially capture information for a wider range of frequencies. It will be beneficial to  understand the impact of signal length, and whether there is an optimal ECG signal length that is able to detect different cardiovascular disorders using different machine learning methods.

\subsection{Utility and performance}
Machine learning  models for ECG classification are ultimately being developed to support diagnosis and hopefully lead to improved patient outcomes. Models that use ECG image transforms as opposed to the raw signal are less common \cite{Hong2020} and more complex as they involve both image transformation and CNN training stages which may hinder their utility in a clinical setting. In that setting, methods such as machine learning using ECG intervals or even deep learning applied to the ECG signal itself may be more desirable. However, the work reported here identified possible benefits of the approach and developed some ideas that are more broadly applicable. Firstly, it presents an alternative approach to detecting cardiovascular conditions in an ECG signal that may be more robust to different physiological noise types. Secondly, as seen in Section \ref{sec:Res}\ref{sec:ClassPerf}, different image transforms (i.e. frequency or morphology based) are better suited to classification of different conditions.

In terms of performance F1 scores, while it is difficult to make a head to head comparison of different deep networks' performance, results found here are comparable to results in the literature. One model for detecting AF from ambulatory ECG had an F1 score of 0.83; ambulatory ECG is more likely to have noise, but the paper does not comment on the impact of this \cite{Hannun2019}. For the scalogram images, the best F1 scores for Normal and AF were 0.75 and 0.94, respectively. Another model used a 1D CNN to classify ECG signals and gave F1 scores for Normal and AF of 0.92 and 0.81 respectively \cite{Goodfellow2018a}, although two things should be noted here. Firstly, the AF dataset was almost seven times smaller than the Normal data which will reduce classification performance for AF \cite{Johnson2019}. Secondly, the ECG data was collected using a handheld device as opposed to the 12-lead data used in the present study. The device used to record the data should be taken into account when interpreting classification results.

On a similar theme, the lead chosen to generate the ECG image transforms (lead II) contributed to the classification performance for the different classes. For example, while the raised heart rate (\(>100\) bpm) characteristic of AF can be seen on any of the 12 leads, the first chest lead is often the best lead to spot the irregular fibrillatory waves \cite{Goldberger2018}. In addition, while STD is often present in multiple leads, it is not always present in lead II \cite{Goldberger2018} and it would be beneficial to evaluate the classification performance of different ECG leads. The ECG preprocessing may have altered ST changes on the ECG to some extent \cite{Lenis2017} which would also contribute to the low F1 scores for STD.

Before any model of this sort could be reliably implemented in the clinic, a better understanding of how it has chosen to classify signals is required. Interpretability tools such as LIME \cite{Ribeiro2016} and layer-wise relevance propagation \cite{Bach2015} are being implemented to understand how these models classified the ECG image transforms, which will additionally help to clarify which regions of the image (i.e. a particular frequency or morphological feature) were most helpful in assigning a particular pathology class label.

\subsection{Limitations}
Regarding the generation of the noisy ECG datasets, it should be noted that filtering can perform differently on healthy or pathological ECG signals \cite{Yang2013}, therefore there may have been a different amount of noise on the Normal, AF and STD signals even before the new physiological noise was applied. In addition, the paper describing the generation of the physiological noise used noted that it was almost impossible to record motion artefact without baseline wander \cite{Moody1984}, suggesting that the results on electrode movement or motion artefact alone are less meaningful. 

There is controversy around the use of ECG transforms generated using the wavelet transform (including scalograms), due to varying frequency resolutions, meaning patterns within an individual subject are identified rather than between subjects \cite{Hong2020}. Future work to evaluate the impact of normalisation and scalogram limit fixing will be beneficial in this area.

Finally, transfer learning models are very sensitive to the choice of network hyperparameters such as the learning rate and number of training epochs, therefore it is unwise to take the values of the performance metrics as definitive. However, these results give insight into how robust these networks are when ECG data with physiological noise is classified. Future work is required to investigate how different network and dataset combinations perform in the presence of ECG noise. There was a slight class imbalance which may have impacted performance, augmented datasets or a custom loss function could address this imbalance in future work.

\section{Conclusion}
We have shown that ECG image transform classification using a CNN is impacted by physiological noise, in particular by electrode movement and motion artefacts. We have also found that different image transform methods may be more appropriate for classifying different cardiovascular conditions, for example a frequency based image for an ECG frequency based condition. With regards to robustness, we found that a network trained on clean data finds it harder to classify noisy data than a network trained on noisy data does to classify clean data. This aligns well with empirical findings that deep network performance is improved by the addition of noise to the training data. Finally, when the future reliability and robustness of a machine learning model used to detect cardiovascular conditions in ECG signals is being evaluated, attention should be paid to level and type of noise in the data used to train the network.

\enlargethispage{20pt}


\aucontribute{JV conceived of and designed the study, ran transfer learning and prepared the first draft of the manuscript. PJA generated the noisy signals and ECG image datasets and edited the manuscript. AS provided input for transfer learning. JV, PJA, PMH and AS developed the noise model. PJA, PMH and NS were involved in funding acquisition. All authors provided input on the shape of the study and the manuscript. All authors read and approved the manuscript for publication.}

\competing{Philip Aston has a patent WO2015121679A1 ``Delay coordinate analysis of periodic data'' which covers the foundations of the SPAR method used in this paper.}

\funding{This project 18HLT07 MedalCare has received funding from the EMPIR programme co-financed by the Participating States and from the European Union’s Horizon 2020 research and innovation programme.}

\ack{Thanks to Claudia Nagel from Karlsruhe Institute of Technology for advice on the ECGdeli software and help obtaining the Lund noise model code. Thanks to the University of Lund for permission to use the noise model and for providing the code. Thanks to Spencer Thomas from the National Physical Laboratory for advice on transfer learning. Thanks to Frederic Brochu from the National Physical Laboratory for advice on parallelising the code and HPC guidance.}



\begin{thebibliography}{99}

\bibitem{Hong2020}
Hong S, Zhou Y, Shang J, Xiao C, Sun J. 2020.
Opportunities and challenges of deep learning methods for  electrocardiogram data: A systematic review.
{\em Comp. Biol. Med.} {\bf 122}, 103801.

\bibitem{Luo2010}
Luo S, Johnston P. 2010.
A review of electrocardiogram filtering.
\textit{J. Electrocardiol.} \textbf{43}.

\bibitem{Ribeiro2020}
Ribeiro AH, Ribeiro MH, Paix\~{a}o GMM, Oliveira DM, Gomes PR, Canazart JA, Ferreira MPS, Andersson CR, Macfarlane PW, Meira Jr W, Sch\"{o}n TB, Ribeiro ALP. 2020.
Automatic diagnosis of the 12-lead ECG using a deep neural network.
\textit{Nat. Commun.} \textbf{11}.

\bibitem{ImageNet}
\textit{ImageNet}.
\href{http://www.image-net.org}{http://www.image-net.org}

\bibitem{Lyon2017}
Aurore L, Minchol\'{e} A, Mart\'{i}nez JP, Laguna P, Rodriguez B. 2017.
Computational techniques for ECG analysis and interpretation in light of their contribution to medical advances.
\textit{J. R. Soc. Interface} \textbf{15}, 20170821.

\bibitem{Deng2009}
Deng J, Dong W, Socher R, Li L-J, Li K, Fei-Fei L. 2009. ImageNet: A large-scale hierarchical image database.
In \textit{2009 IEEE conference on computer vision and pattern recognition.}, 248--255.

\bibitem{Byeon2019}
Byeon YH, Pan SB, Kwak KC. 2019.
Intelligent deep models based on scalograms of electrocardiogram signals for biometrics.
{\em Sensors} {\bf 19}, 935.

\bibitem{Sun2019}
Sun W, Zeng N, He Y. 2019.
Morphological arrhythmia automated diagnosis method using gray-level co-occurrence matrix enhanced convolutional neural network.
\textit{IEEE ACCESS} \textbf{7}.

\bibitem{Mathunjwa2021}
Mathunjwa BM, Lin Y-T, Lin C-H, Abbod MF, Shieh J-S. 2021.
ECG arrhythmia classification by using a recurrence plot and convolutional neural network.
\textit{Biomed. Signal Proces.} \textbf{64}.

\bibitem{Li2018}
Li Y, Zhang Y, Zhao L, Zhang Y, Liu C, Zhang L, Zhang L, Li Zhensheng, Wang B, Ng E, Li J, He Z. 2018.
Combining convolutional neural network and distance distribution matrix for identification of congestive heart failure.
\textit{IEEE ACCESS} \textbf{6}.

\bibitem{Aston2019}
Aston PJ, Lyle JV, Bonet-Luz E, Huang CLH, Zhang Y, Jeevaratnam K, Nandi M. 2019.
Deep learning applied to attractor images derived from ECG signals for detection of genetic mutation.
{\em Comp. Cardiol.} {\bf 45}, 097.

\bibitem{PerezAlday2020}
Perez Alday EA, Gu A, Shah A, Liu C, Sharma A, Seyedi S, Bahrami Rad A, Reyna M, Clifford G. 2020.
Classification of 12-lead ECGs: the PhysioNet - Computing in Cardiology Challenge 2020 (version 1.0.1).
\textit{PhysioNet}.

\bibitem{Goldberger2018}
Goldberger AL, Goldberger ZD, Shvilkin A. 2018.
{\em {Goldberger's Clinical Electrophysiology: A Simplified Approach}}.
Elsevier Inc., 9th edition.

\bibitem{Pollehn2002}
Pollehn T, Brady WJ, Perron AD, Morris F. 2002.
The electrocardiographic differential diagnosis of ST segment depression.
\textit{Emerg. Med. J.} \textbf{19}, 129--035.

\bibitem{Pilia2020}
Pilia N, Nagel C, Lenis G, Becker S, D\"{o}ssel, Loewe A. 2020.
ECGdeli - An open source ECG delineation toolbox for MATLAB.
\href{https://github.com/KIT-IBT/ECGdeli}{https://github.com/KIT-IBT/ECGdeli}

\bibitem{Moody1984}
Moody GB, Muldrow WK, Mark RG. 1984.
A noise stress test for arrhythmia detectors.
\textit{Comput. Cardiol.} \textbf{11}.

\bibitem{Goldberger2003}
Goldberger AL, Amaral LAN, Glass L, Hausdorff JM, Ivanov PC, Mark RG, Mietus JE, Moody GB, Peng CK, Stanley HE. 2003.
PhysioBank, PhysioToolkit, and PhysioNet: Components of a new  research resource for complex physiologic signals.
{\em Circ.}, {\bf 101}, e215--e220.

\bibitem{Petrenas2017}
Petr\.enas A, Marozas V, Solo\u{s}enko A, Kubilius R, Skibarkien\.e J, Oster J, S\"{o}rnmo L. 2017.
Electrocardiogram modeling during paroxysmal atrial fibrillation: application to the detection of brief episodes.
\textit{Physiol. Meas.} \textbf{38}.

\bibitem{Aston2018}
Aston PJ, Christie MI, Huang YH, Nandi M. 2018.
Beyond HRV: Attractor reconstruction using the entire cardiovascular waveform data for novel feature extraction.
{\em Phys. Meas.} {\bf 39}.

\bibitem{Nandi2018}
Nandi M, Venton J, Aston PJ. 2018.
A novel method to quantify arterial pulse waveform morphology: Attractor reconstruction for physiologists and clinicians.
\textit{Physiol. Meas.} \textbf{39}.

\bibitem{Meek2002}
Meek S, Morris F. 2002.
ABC of clinical electrocardiography.Introduction. I-Leads, rate, rhythm, and cardiac axis.
\textit{BMJ (Clinical research ed.)} \textbf{324}, 415--418.

\bibitem{Goodfellow2016}
Goodfellow I, Bengio Y, Courville A. 2016 
\textit{Deep Learning Book}.
MIT Press. See \href{https://www.deeplearningbook.org}{https://www.deeplearningbook.org}

\bibitem{An1996}
An G. 1996.
The effects of adding noise during backpropagation training on a generalisation performance.
\textit{Neural Comput.} \textbf{8}, 643--674.

\bibitem{Strodthoff2020}
Strodthoff N, Wagner P, Schaeffter T, Samek W. 2020.
Deep Learning for ECG Analysis: Benchmarks and Insights from PTB-XL.
\textit{arXiv:2004.13701v1 [cs.LG]}.

\bibitem{Han2020}
Han X., Hu Y, Foschini L, Chinitz L, Jankelson L, Ranganath R. 2020.
Deep learning models for electrocardiograms are susceptible to  adversarial attack.
{\em Nat. Med.} {\bf 26}, 360--363.

\bibitem{Kurakin2017}
Kurakin A, Goodfellow IJ, Bengio S. 2017.
Adversarial machine learning at scale.
\textit{arXiv:1611.01236v2 [cs.CV]}.

\bibitem{Oster2015}
Oster J, Clifford GD. 2015
Impact of the presence of noise on the RR interval-based atrial fibrillation detection.
\textit{J. Electrocardiol.} \textbf{48}, 947--951.

\bibitem{Hannun2019}
Hannun AY, Rajpurkar P, Haghpanahi M, Tison GH, Bourn C, Turakhia MP, Ng AY. 2019.
Cardiologist-level arrhythmia detection and classification in  ambulatory electrocardiograms using a deep neural network.
{\em Nat. Med.} {\bf 25}, 65--69.

\bibitem{Goodfellow2018a}
Goodfellow SD, Goodwin A, Greer R, Laussen PC, Eytan D, Goodfellow SD, Goodwin A, Greer R, Laussen PC, Mazwi M, Eytan D. 2018.
Towards understanding ECG rhythm classification using convolutional  neural networks and attention mappings.
{\em Proc. 3rd ML Healthc. Conf.} {\bf 85}, 83--101.

\bibitem{Johnson2019}
Johnson JM, Khoshgoftaar TM. 2019.
Survey on deep learning with class imbalance.
\textit{J. Big Data} \textbf{6}

\bibitem{Lenis2017}
Lenis G, Pilia N, Loewe A, Schulze WHW, D\"{o}ssel O. 2017.
Comparison of baseline wander removal techniques considering the preservation of ST changes in the ischemic ECG: a simulation study.
\textit{Comput. Math. Method M.} \textbf{2017}, 9295029.

\bibitem{Ribeiro2016}
Ribeiro MT, Singh S, Guestrin C. 2016.
"Why should I trust you?": Explaining the predictions of any classifier.
\textit{Proceedings of the 22nd ACM SIGKDD International Conference on Knowledge Discovery and Data Mining}.

\bibitem{Bach2015}
Bach S, Binder A, Montavon G, Klauschen F, M{\"{u}}ller KR, Samek W. 2015.
On pixel-wise explanations for non-linear classifier decisions by  layer-wise relevance propagation.
{\em PLOS ONE} {\bf 10}, 1--46.

\bibitem{Yang2013}
Yang M, Liu B, Zhao M, Li F, Wang G, Zhou F. 2013.
Normalizing electrocardiograms of both healthy persons \& cardiovascular disease patients for biometric authentication.
\textit{PLOS ONE} \textbf{8}.

\end{thebibliography}
\end{document}